\documentclass{pasj00}

\begin{document}
\SetRunningHead{H. Murakami et al.}{TeV J2032$+$4130}
\Received{2011/06/14} 
\Accepted{2011/07/29} 

\title{Detection of X-Ray Emission from the Unidentified TeV Gamma-Ray Source
TeV J2032$+$4130}


\author{Hiroshi \textsc{Murakami}, Shunji \textsc{Kitamoto}}
\affil{%
Department of Physics, Rikkyo University, 3-34-1 Nishi-Ikebukuro, Toshima-ku, Tokyo 171--8501}
\email{hiro@rikkyo.ac.jp}

\author{Akiko \textsc{Kawachi}}
\affil{%
Department of Physics, Tokai University, Kitakaname 1117, Hiratsuka, Kanagawa 259--1292}

\and

\author{Takeshi \textsc{Nakamori}}
\affil{%
Research Institute for Science and Engineering, Waseda University,
3-4-1 Okubo, Shinjuku-ku, Tokyo 169--8555}

\KeyWords{acceleration of particles --- X-rays: individual (TeV J2032$+$4130)
--- X-rays: ISM --- pulsars: individual (PSR J2032$+$4127)} 

\maketitle

\begin{abstract}
We observed the first unidentified TeV $\gamma$-ray source 
TeV J2032$+$4130 with Suzaku. 
Owing to Suzaku's high sensitivity for detection of diffuse X-ray emission,
we found two
small structures in the TeV emitting region. One of them is coincident with
a $\gamma$-ray pulsar PSR~J2032$+$4127, which was discovered by the 
Fermi Gamma-ray Space Telescope. By subtracting
contribution of point sources estimated by Chandra data, we
obtained diffuse X-ray spectrum. The X-ray spectrum can be reproduced by a
power-law model with a photon index of $\sim$~2, and an X-ray flux of 
$2\times10^{-13}$ erg~s$^{-1}$~cm$^{-2}$. The ratio of the $\gamma$-ray flux to the
X-ray flux is about 10. If the origin of the TeV $\gamma$-ray is inverse
Compton scattering of microwave background by high energy electrons,
the ratio corresponds to the magnetic field strength of $\sim$~1~$\mu$G. 
However, the smaller size of the X-ray emission than that of the TeV emission 
suggests that energy loss of the electrons can explain the large
ratio of the $\gamma$-ray flux with a reasonable magnetic field strength
of a few $\mu$G. 

\end{abstract}

\section{Introduction}
The stereoscopic technique of atmospheric Cerenkov telescopes
improved the angular resolution for detecting TeV $\gamma$-rays
and thus increased the number of TeV $\gamma$-ray sources.
Some of new TeV objects have no counterparts at other wavelengths and
are called unidentified TeV $\gamma$-ray objects.
These objects provide key information in the investigation of
the origin of high-energy cosmic rays. Multi-wavelength
observations of these objects 
are very important in elucidating the emission mechanism and 
successful identification of them.

TeV J2032+4130 was the first unidentified TeV $\gamma$-ray source discovered by
HEGRA (\cite{Aha2002}). The TeV emission 
exhibits a significant extension 
with a radius of \timeform{6.2'} and a center of gravity of (R.A., Dec) = 
(\timeform{20h31m57.0s}, \timeform{41D29'56.8''}) \citep{Aha2005a}.
The position is coincident with an OB association, Cyg OB2, and located
in north of a microquasar, Cyg X-3. 
These two sources had been suggested to be
possible origins of TeV $\gamma$-rays, but no firm evidence to determine
the counterpart has been found.
TeV $\gamma$-ray emission of this region has also been reported by
other telescopes: Whipple, MAGIC, and Milagro (\cite{Konopelko2007},
\cite{Alb2008}, \cite{Abdo2007}).

In the X-ray band, Chandra and XMM-Newton observed the TeV~J2032+4130 region.
The first observation by Chandra detected 27 point sources within the observed
field ($\sim$ \timeform{17'}) with the exposure time of 5 ks (\cite{MHG2003}, MHG2003 hereafter,).
\citet{But2003} also resolve 19 point sources
above the threshold of 2.5 $\sigma$ by adapting the $\tt wavdetect$ tool
to the same data.
Then follow-up deeper 50~ks observation remarkably increased
the number of detected sources: $240$ sources
in almost the same filed (\cite{But2006}, But2006 hereafter). 

XMM-Newton also detected many point sources in the wider FOV ($\sim$ \timeform{30'}).
By subtracting the contribution of detected point sources, \citet{Horns2007}
indicates a hint of diffuse emission
extending about the size of TeV emission region.
However, this result may still include
contribution of faint point sources which were resolved by Chandra,
due to the moderate angular resolution of XMM-Newton.

The detection of a diffuse X-ray emission for the 50~ks Chandra
data has been also reported
by \citet{MGH2007} with the same analysis technique. However, 
the spectral model can not be constrained because of the low photon
statistics observation by Chandra.

Recently, the Large Area Telescope on the Fermi Gamma-ray
Space Telescope detected $\gamma$-rays from this region, with the energy from 
20~MeV to 300~GeV (1FGL J2032.2+4127, \cite{Abdo2010a}). 
In addition, this $\gamma$-ray source showed a pulsation
with a pulse period of 143~ms (PSR~J2032$+$4127; \cite{Abdo2009}, 2010b). 
Subsequent observation of a radio band also detected a pulsation
with the consistent position and the pulse frequency of the $\gamma$-ray pulsar.
The position of the pulsar is 
coincident with the optical point source of the number 213 in 
\citet{MT91}
(MT91 213; \cite{Camilo2009}). 

These results imply that the origin of TeV $\gamma$-rays also relates to this
$\gamma$-ray pulsar. Active pulsars are losing a significant part of the 
energy via relativistic particles, and forms pulsar wind nebulae (PWNe).
PWNe emit synchrotron radiation from radio to X-ray bands.
In addition, some PWNe are
found to be TeV emitter (\cite{Gaensler2006}, \cite{Kar2010}). 
Thus PSR~J2032$+$4127 is a possible candidate for the counterpart of
TeV~J2032+4120. 

The distance to the PSR~J2032$+$4127 is estimated to be
3.6~kpc by measuring the dispersion measure in 
the radio band \citep{Camilo2009}. While, the distance to
the Cygnus OB2 is estimated to be 1.7~kpc by a spectroscopic
observation of OB stars \citep{Hanson2003}. 
We adopt former value as the distance to X-ray emission.

In this paper, we report X-ray observation of TeV J2032+4130
with Suzaku, which has a higher sensitivity for detecting 
diffuse X-ray emission
with the large effective area and the low stable background. 
We analyze the diffuse X-ray spectrum of the PWN in detail.

Though Suzaku has the advantage in detecting diffuse X-ray sensitivity,
the angular resolution is not sufficient to resolve point sources.
To properly estimate the contribution of point sources,
we also reanalyze the Chandra data. There are many point 
sources in and near the Cygnus OB2 region, which are one of the candidate for
the origin of TeV emission (e.g., \cite{But2006}). We resolve point sources
within strict parameters, and subtract the point source flux
from the diffuse emission. Thus we investigate diffuse emission
by combination of Suzaku and Chandra.

\section{Observations}
\subsection{Suzaku}
We observed the TeV~J2032$+$4130 region with Suzaku \citep{Mitsuda2007}
on December 17 and 18, 2007. Suzaku has a moderate angular resolution
and a large effective area. This characteristic is suitable for
detecting weak diffuse emission.
The observations were made using three CCD
cameras (X-ray Imaging Spectrometer, XIS; \cite{Koyama2007}) on the
focal planes of the X-Ray Telescopes (XRT; \cite{Ser2007}).
One of the cameras (XIS1) has a back-illuminated (BI) CCD, and the
others (XIS0, 3) contain front-illuminated (FI) CCDs.
Each of the XIS sensors was operated in the normal clocking mode
with the 5$\times$5 or 3$\times$3 editing mode.

We used clean events processed with the pipeline version of 2.1.6.16.
Data taken during the passage through the South Atlantic Anomaly,
at elevation angles less than 5$^\circ$ from the night Earth rim,
or 20$^\circ$ from the day Earth rim
were excluded. After this filtering, the net observing time was
about 40~ks. 

\subsection{Chandra}
Chandra observed TeV~J2032$+$4130 region twice: 
an earlier short observation (August 11 2002, obsid=4358) and a deep follow-up
observation (July 19 2004, obsid=4501).
The exposure times were 5~ks and $\sim$~49~ks, respectively.
Chandra has a superior angular resolving capability.
We analyzed Chandra data in order to estimate the
contribution of point sources,
using the Chandra Interactive
Analysis of Observations (CIAO) software version 4.0.2 with CALDB
version 3.4.3. 
The detailed results about the Chandra observations have already
been reported in \citet{But2006}.

\section{Results}
\subsection{Images}
\begin{figure}
  \begin{center}
    \FigureFile(80mm,80mm){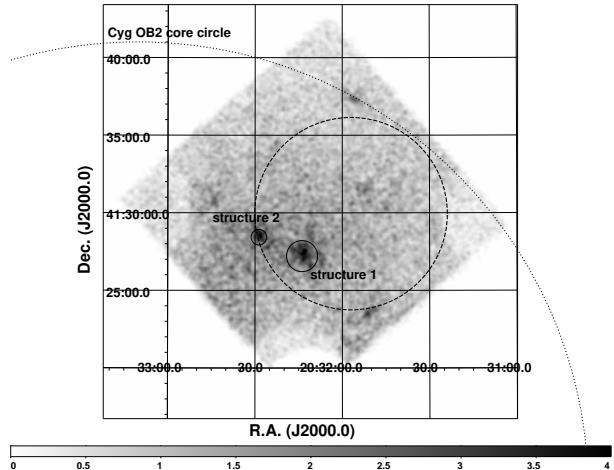}
  \end{center}
  \caption{Suzaku image of TeV J2032$+$4130 (0.5--10.0 keV).
    Dashed circle indicates the region of diffuse TeV emission. X-ray structures are shown
    in solid circles. Dotted circle is core circle of Cyg OB2 \citep{Kno2000}.}
  \label{fig:suzaku_image}
\end{figure}

We first construct a total energy band image of the TeV~J2032$+$4130 region
with Suzaku data.
Fig.~\ref{fig:suzaku_image} shows an XIS image of 0.5--10.0 keV band. All
three CCD data are combined. A dashed circle indicates the TeV
diffuse emission region \citep{Aha2005a}.

There are two diffuse structures in the circle (structures 1 and 2 in Fig.~1).
The extent of X-ray emission is significantly larger than the point spread
function.
Assuming a Gaussian profile, their sizes are $\sim$~\timeform{1.1'} 
for both structures;
however, they are much smaller than the TeV emission.

These X-ray emitting structures are located at the south eastern part of the
TeV $\gamma$-ray region, and included in the OB star association Cyg OB2
(dotted circle in Fig~\ref{fig:suzaku_image}).
There must be
many point sources in the field. We
estimate the contribution of the point sources by the Chandra
deep exposure data, which has superior angular resolution,
and can resolve weak point sources.

We extracted point sources by the CIAO ``$\tt wavdetect$'' software of a
wavelet method \citep{Freeman02a}. The threshold significances of
$\tt wavdetect$ were set at $10^{-6}$ for the source list
and at 0.001 for
the background estimation. The wavelet scales were
1, $\sqrt{2}$, 2, $2\sqrt{2}$, 4, $4\sqrt{2}$, 8, $8\sqrt{2}$, and 16
pixels. We then resolved 254 point sources from the whole region 
of ACIS-I CCDs.
The structure 1 region includes 8 sources, while the
structure 2 region includes 2 sources.
158 sources are located in the TeV $\gamma$-ray region (dashed circle in
Fig.~\ref{fig:suzaku_image}), excluding structures 1 and 2.

\begin{figure}[htb]
  \begin{center}
    \FigureFile(77mm,77mm){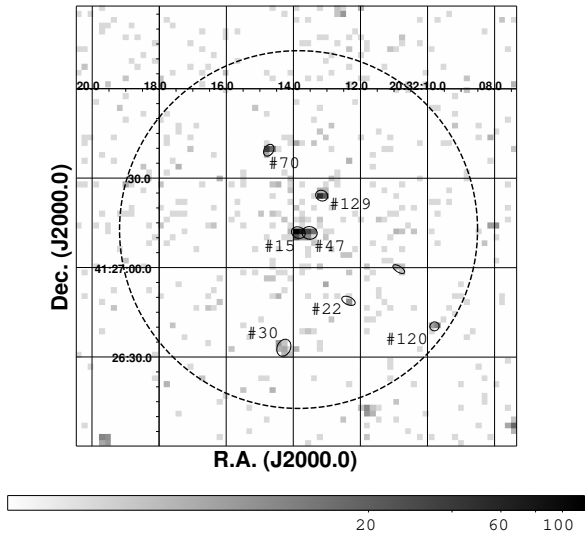}
  \end{center}
  \caption{Chandra image of structure 1 (dashed circle). 
Point sources are indicated by solid ellipses. 
Labels show the source numbers in Butt et al. (2006). \#70 is 
coincident with \#17 in MHG2003, while the combination of
\#15 and \#47 would be \#18 in MHG2003 (see also Table~\ref{tab:count}).
A newly detected source has no number. \#129 is
coincident with the $\gamma$-ray pulsar PSR~J2032+4127.}
  \label{fig:cxo_image}
\end{figure}

Fig.~\ref{fig:cxo_image} shows a Chandra image around structure 1.
Point sources are indicated by solid ellipses with the source
numbers in Butt et al. (2006). A point source without a number
is newly resolved by our analysis.
One of the point sources in structure~1, \#129,
is coincident with the $\gamma$-ray pulsar discovered by Fermi (Camilo
et al. 2009).

\subsection{Spectra}
First, we constructed a spectrum of point sources by Chandra
observation in order to estimate the contribution of point sources
in the Suzaku image.
We collected all the events from the point sources in structure 1, 2
and the remainder of the TeV $\gamma$-ray emitting region.
X-ray photons are extracted from an ellipse with the axes of 3~$\sigma$
of the 2-D Gaussian calculated by $\tt wavdetect$ for each source.
To reproduce
spectra, we fit the spectrum using a phenomenological model of 
a power-law with an interstellar absorption. 
The best-fit parameters are shown in ``point sources'' rows in
Table~\ref{tab:spec}.

\begin{figure}[htp]
  \begin{center}
    \FigureFile(60mm,50mm){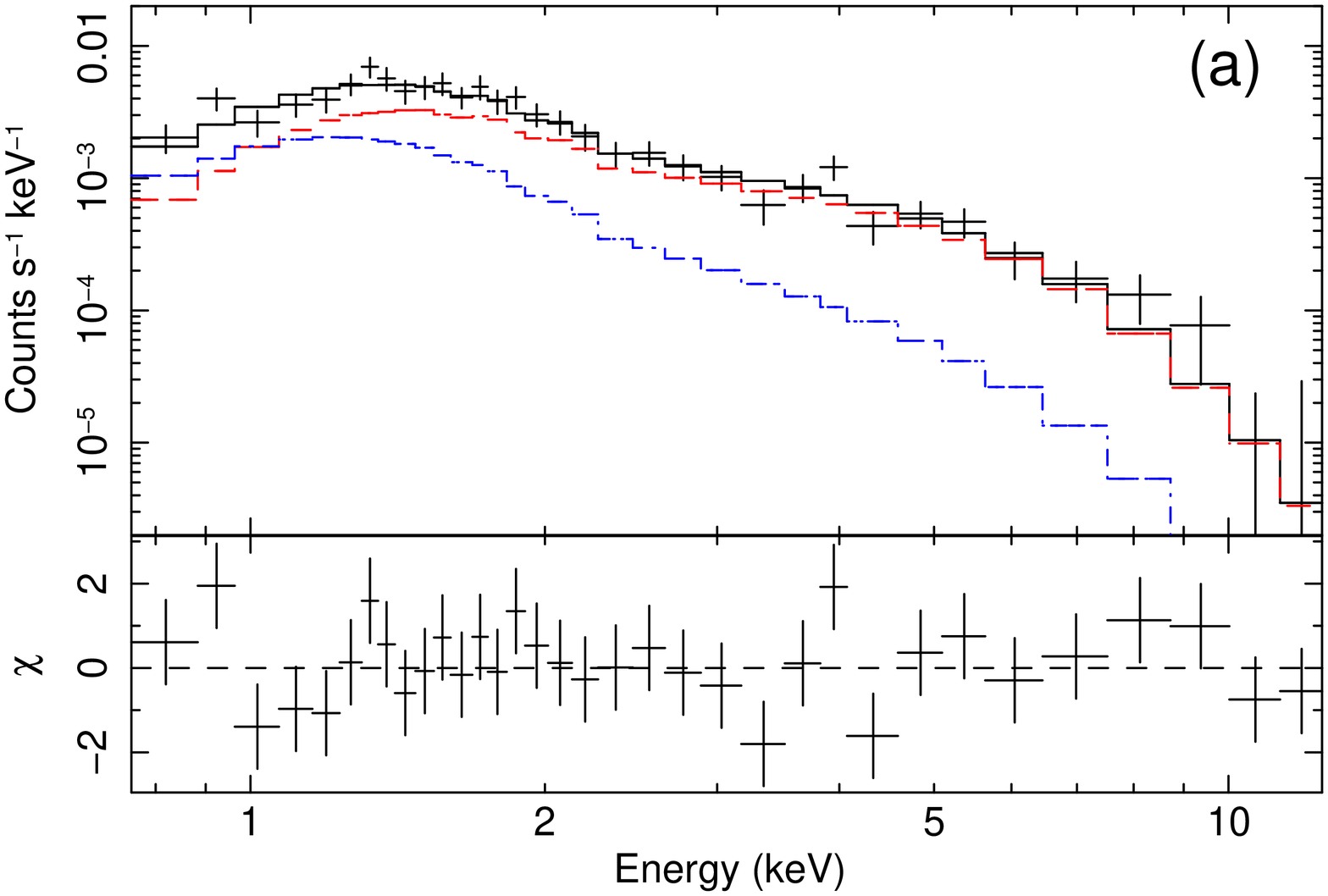}
    \FigureFile(60mm,50mm){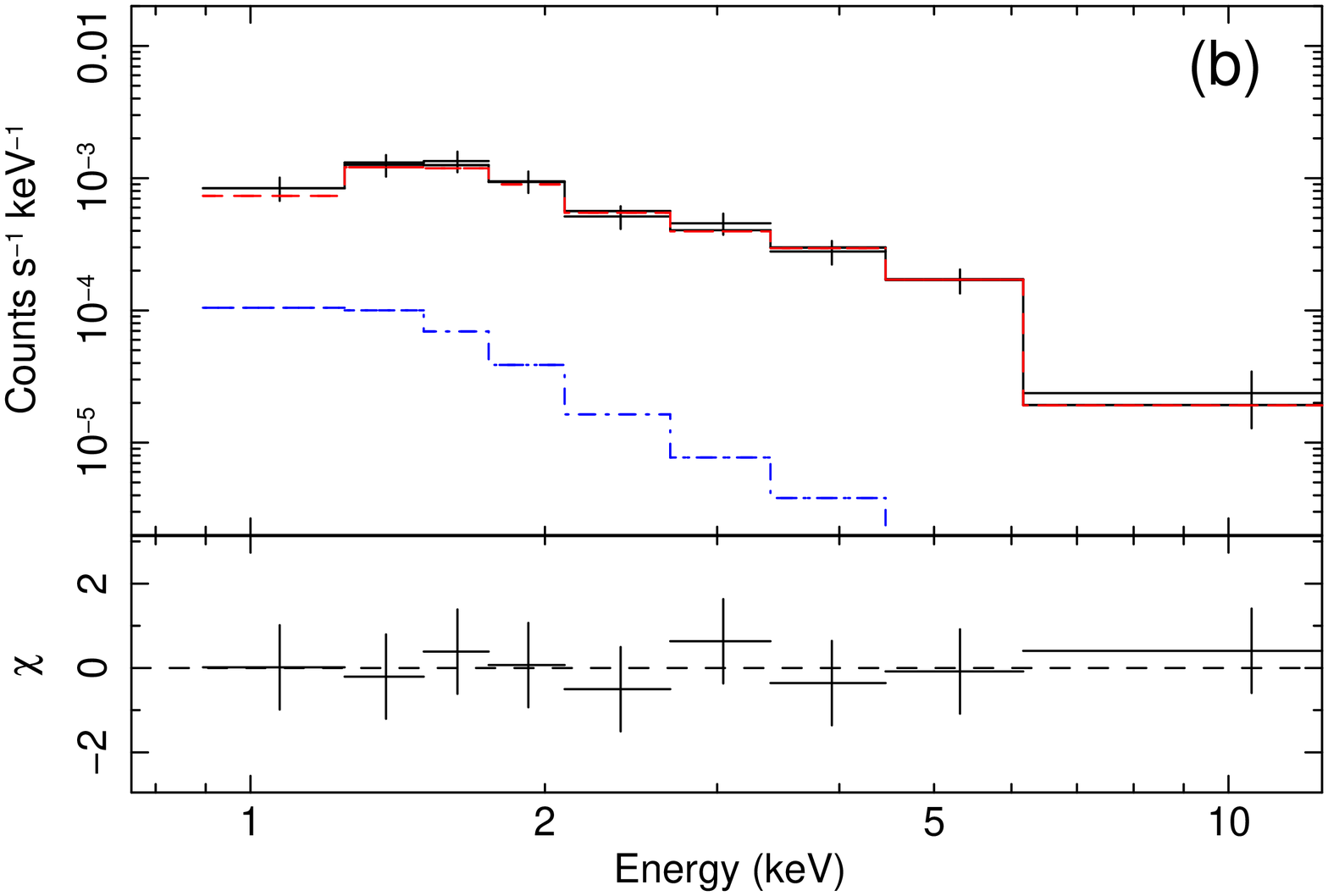}
    \FigureFile(60mm,50mm){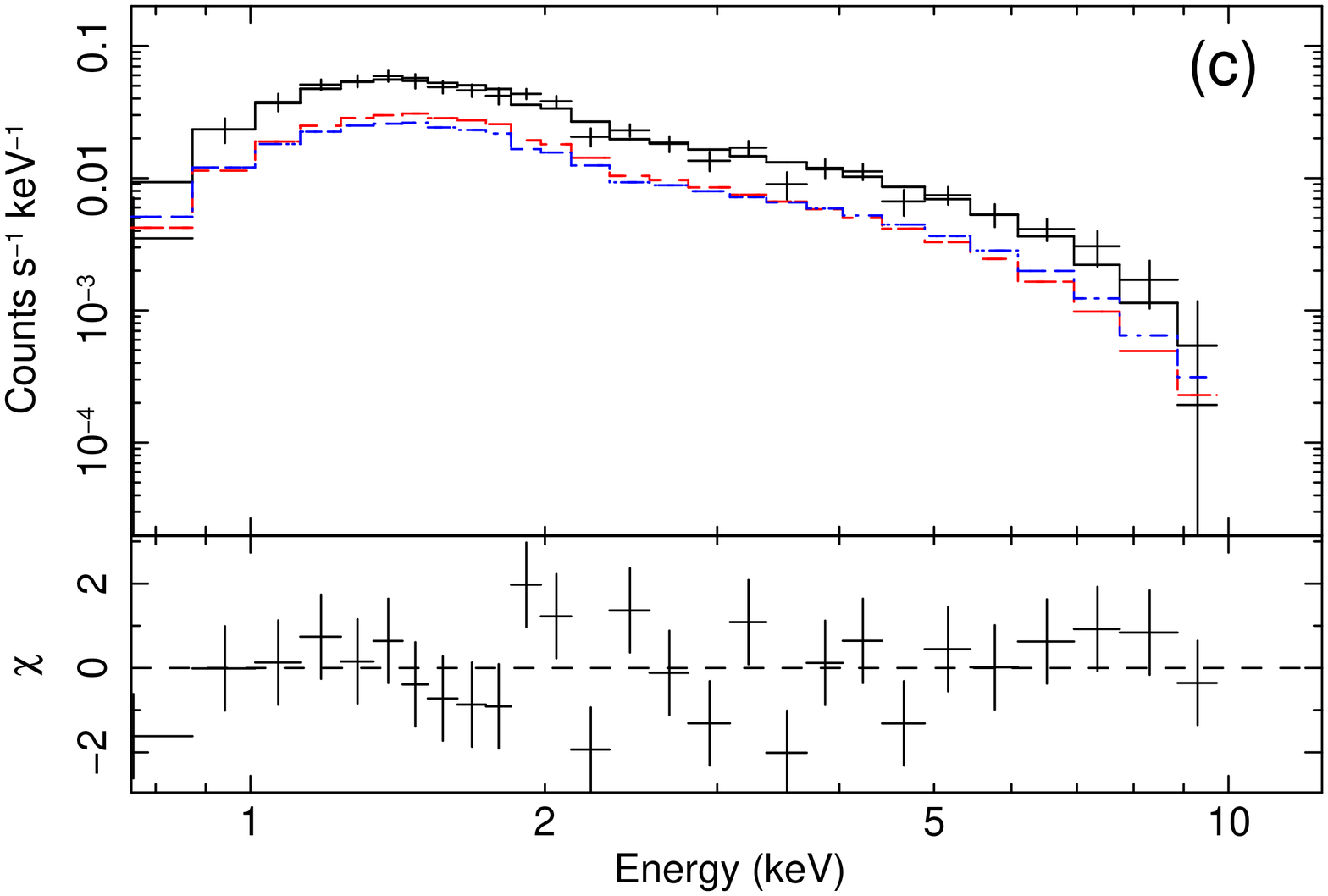}
  \end{center}
  \caption{Suzaku spectrum of structure 1, structure 2
and the remaining region of the TeV emission (a, b, and c, respectively). 
Contribution of point sources is
indicated by the blue dotted line in each panel.
Red dashed line shows diffuse component.}
\label{fig:suzaku_spec}
\end{figure}

Then we constructed a spectrum of diffuse emission using Suzaku data.
The extracted spectrum of structure~1 is shown in
Fig.~\ref{fig:suzaku_spec}(a). The spectra and responses of three CCDs 
are combined.
We include the contribution of the point sources by adding its
best-fit model into the model spectrum, which is
indicated by the blue dotted line in
Fig.~\ref{fig:suzaku_spec}(a). 
Thus, we obtained best-fit parameters of diffuse X-ray emission
for structure~1 (Table~\ref{tab:spec}). The spectrum can be reproduced 
by an absorbed power-law with a photon index ($\Gamma$) of 2.1, and an absorption 
column ($N_{\rm H}$) of 0.6 $\times 10^{22}$ cm$^{-2}$. 
The X-ray flux is about $2.0 \times 10^{-13}$~erg~s$^{-1}$~cm$^{-2}$
(2.0--10.0 keV). 

We also derived best fit parameters of X-ray spectra extracted
from structure 2 and from the remainder in the same manner
(Table~\ref{tab:spec}). 
The best fit models of diffuse and point sources are
plotted in Fig.~\ref{fig:suzaku_spec}(b) and (c).
The absorption corrected luminosities (2--10 keV) of diffuse components 
are $3.1 \times 10^{32}$~erg~s$^{-1}$, $3.0 \times 10^{32}$~erg~s$^{-1}$,
 and $14 \times 10^{32}$~erg~s$^{-1}$, for structures  1, 2 and 
the remaining region, respectively.

In structures 1 and 2, the spectra of point sources are softer
than the diffuse component. It indicates that the point source
rejection method properly works. It also implies that the origin
of diffuse X-ray emission is not a concentration of 
faint point sources. Most of the point sources in 
structure 1 and 2 are stars, which generally exhibit softer
X-ray emission. 
The typical temperature of the Cygnus OB2 stars is 1.35 keV \citep{Alba2007}.
Indeed, the spectra of point sources in structure 1 and 2 can be represented
by thin thermal plasma model with the temperature of $\sim 1$~keV.

\begin{table*}[hbt]
  \caption{Best-fit parameters of X-ray spectra$^\ast$.}
  \label{tab:spec}
  \begin{center}
    \begin{tabular}{ccccccc}
      \hline\hline\\[-6pt]
      && $N_{\rm H}$ & $\Gamma$ & $F_{\rm X}^\dagger$ & $L_{\rm X}^\ddagger$ & $\chi^2$ (d.o.f) \\
      && {\footnotesize ($10^{22}$~cm$^{-2}$)} & & {\footnotesize ($10^{-13}$~erg~s$^{-1}$~cm$^{-2}$)} 
      & {\footnotesize ($10^{32}$~erg~s$^{-1}$)} & \\
      \hline
structure 1&      point sources & 0.4$^{+0.2}_{-0.1}$ & 3.1$^{+0.5}_{-0.4}$ & 0.32$^{+0.03}_{-0.02}$ & & 23.0 (14) \\
      &diffuse & 0.6$^{+0.4}_{-0.2}$ & 2.1$^{+0.4}_{-0.3}$ & 2.0$^{+0.1}_{-0.2}$ &3.1  & 28.6 (31) \\
structure 2&      point sources & 0.4$^{+0.8}_{-0.4}$ & 3.5$^{+2.6}_{-0.9}$ & 0.025$\pm 0.006$ & & 0.70 (3) \\
      &diffuse & 0.6$^{+0.5}_{-0.3}$ & 1.9$^{+0.5}_{-0.2}$ & 1.9$^{+0.3}_{-0.2}$ & 3.0  & 1.15 (6) \\
The remain&      point sources & 0.47$\pm 0.07$ & 1.8$\pm 0.1$ & 9.8$^{+0.2}_{-0.3}$ & & 47.1 (23) \\
      &diffuse & 0.7$^{+0.2}_{-0.3}$ & 2.2$^{+0.4}_{-0.3}$ & 9.0$^{+0.8}_{-0.7}$ & 14 & 28.3 (24) \\
      \hline
      \multicolumn{6}{l}{\footnotesize $^\ast$: The uncertainties are 90 \% confidence level.}  & \\
      \multicolumn{6}{l}{\footnotesize $^\dagger$: X-ray flux in the 2--10 keV band} & \\
      \multicolumn{6}{l}{\footnotesize $^{\ddagger}$: Absorption corrected X-ray luminosity in the 2--10 keV band. 
                         The distance is assumed to be 3.6~kpc.} & \\
    \end{tabular}
  \end{center}
\end{table*}

\subsection{Time variation of point sources}
We estimated the contribution of point sources using the Chandra data.
However, there is uncertainty in the estimated flux because of
a possible long term variability of 
sources. Indeed, \citet{MGH2007} reported a detection of
transient X-ray sources in the TeV emission region. 
We evaluate the uncertainty from the two Chandra observations of TeV~J2032+4130 region: 
an initial 5~ks observation in 2002 (MHG2003) and a deep follow-up 
exposure of 50~ks in 2004 (But2006).
We compared the X-ray count rate of point sources between
these two observations. 

Earlier observation has poor photon statistics, and detected only 2 sources 
inside of structure~1. These sources are the \#17 and \#18 in MHG2003. 
The follow-up observation detected 7 point sources within the same field
as shown in Fig.~\ref{fig:cxo_image}.
We made the identification between two observations by the coordinates of these sources.
\#70 in But2006 is coincident with MHG2003 \#17. These sources are identical.
While, there are two sources at the position of MHG2003 \#18: \#15 and \#47 in But2006.
These sources were too close to resolve in the earlier observation. We consider
the count rate of \#18 in MHG2003 to be the total of these two sources.
Structure 2 only includes identification \#14 in 
MHG2003. This source is coincident with 
\#150 in But2006.
We have summarized the count rates of these sources in Table~\ref{tab:count}.

The count rate of MHG2003 \#17 is 3.0~$\times 10^{-3}$~counts~s$^{-1}$ in
2002, and 1.5~$\times 10^{-3}$~counts~s$^{-1}$ in 2004. The time
variation is $\sim$2.  For MHG2003 \#18, the count rate is 7.0~$\times
10^{-3}$~counts~s$^{-1}$ in 2002, which is similar to 7.2~$\times
10^{-3}$~counts~s$^{-1}$, the combined count rate of \#15 and \#47
in \citet{But2006}.

The count rate of MHG2003 \#14 in structure 2 is
$\sim 4.4~\times 10^{-3}$~counts~s$^{-1}$ in 2002 
and $0.62~\times 10^{-3}$~counts~s$^{-1}$ in 2004.
The point source indicates the time variability as large as a factor of 7.

As shown in Table~\ref{tab:spec}, the contribution of point sources
is about 16.5\% and 1.3\% for
structures 1 and 2, respectively. Even in the largest case of the time 
variety obtained above (factors 2 and 7), 
the contribution is about 33\% and 9\%, respectively.
Although these uncertainties of point source fluxes have an influence
on the spectrum analysis of diffuse emission, the diffuse emission
cannot be explained by the point sources.

\begin{table*}[htb]
  \caption{Comparison of X-ray count rate of point sources derived from Chandra data.
The time variations of MHG2003 17 and 14 are $\sim 2$ and $\sim 7$, respectively.
MHG2003 18 is considered to be a combination of two point sources 15 and 47 in But2006.}
  \label{tab:count}
  \begin{center}
    \begin{tabular}{cccccc}
      \hline\hline\\[-6pt]
      &\multicolumn{2}{c}{identification} & &\multicolumn{2}{c}{count rate}\\
      &&&& \multicolumn{2}{c}{($\times 10^{-3}$ count/sec)} \\
      \cline{2-3} \cline{5-6}\\[-6pt]
      &MHG2003 & But2006 && 2002 & 2004 \\
      \hline

            & \#17      &  \#70    & & 3.0   & 1.5 \\
structure 1 & \#18      &  \#15    & & 7.0   & 4.8 \\
            &         &  \#47    & &      &  2.4 \\
\hline
structure 2 & \#14      &  \#150   & & 4.4   & 0.62 \\
      \hline
    \end{tabular}
  \end{center}
\end{table*}

\section{Discussion}

We obtained X-ray spectrum of diffuse emission around the $\gamma$-ray pulsar
PSR~J2032$+$4127 (structure~1). 
The photon index is determined to be $\sim$~2.
This value is coincident with the typical index of X-ray spectrum
from PWN: $\Gamma~\simeq$~1--2 \citep{Kar2008}. 
In this section, we consider that the diffuse X-ray emission 
is radiated by the pulsar wind nebula, and discuss the radiation mechanism.

\subsection{Energy injection}
First, we discuss the energy injection rate. The spin-down power of the
$\gamma$-ray pulsar, PSR~J2032$+$4127,
is calculated to be about $2.63\times10^{35}$~erg~s$^{-1}$
(\cite{Abdo2010b}). 
Meanwhile, the isotropic luminosity of the GeV $\gamma$-ray
pulsar is $1.4~\times~10^{35}~(d/3.6 {\rm kpc})^{2}$~erg~s$^{-1}$
(\cite{Camilo2009}), where $d$ is the distance to the pulsar.
At most about a half of the energy is emitted as GeV $\gamma$-rays with an
assumption of isotropic radiation. The ratio can be smaller in the case of
collimated radiation. 
The intensity of off-pulse emission is almost the same as background level,
which is estimated from surrounding annulus region of the pulsar
(Fig. A-42 in \cite{Abdo2010b}). Indeed, detailed analysis of off-pulse 
spectrum cannot constrain the flux level \citep{Ack2011}.
A large portion of GeV $\gamma$-rays originate from the pulsar's 
magnetosphere, and the luminosity of PWN is negligible in this energy band.
Consequently, we cannot obtain meaningful GeV $\gamma$-ray flux for PWN.
We take no account of the GeV $\gamma$-ray band for 
discussing the energy injection rate from the pulsar.

X-ray luminosity and TeV $\gamma$-ray luminosity,
which are considered to be diffuse emission, are
$\sim 3\times10^{32}$~erg~s$^{-1}$ and $\sim 2\times10^{33}$~erg~s$^{-1}$,
respectively (1.0--10 TeV; \cite{Aha2005a}, \cite{Alb2008}).
These are two or three orders of magnitude smaller than the spin-down energy.
The spin-down energy is enough to supply X-ray and TeV $\gamma$-ray emission.
Therefore the energy of the diffuse emission 
in TeV and X-ray could be supplied by the pulsar, as a PWN.
The ratio of $\sim 10^{-3}$ between X-ray luminosity and the spin-down energy of 
the pulsar is typical for X-ray emission of PWN (e.g., Figure~10 in \cite{Kar2009}).

\subsection{Emission mechanism}
\begin{figure}
  \begin{center}
    \FigureFile(70mm,70mm){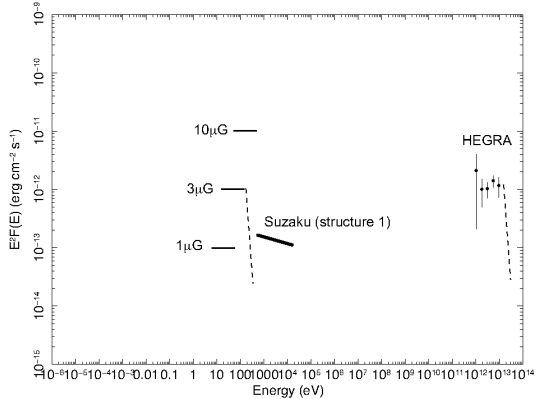}
  \end{center}
  \caption{Multiband spectrum of
TeV J2032$+$4130 (TeV: HEGRA spectrum in \cite{Aha2005a}, X-ray: Best-fit
model of the diffuse component of structure~1).
Solid lines show synchrotron radiation models with various magnetic field
estimated from TeV flux.
Dashed line shows an energy cutoff model.}
\label{fig:wide_spec}
\end{figure}

In Fig.~\ref{fig:wide_spec}, we plot the fluxes of X-rays and 
TeV $\gamma$-rays. The X-ray flux is indicated by best-fit model of
the diffuse component spectrum of structure~1.
The ratio of flux between TeV and the X-ray bands 
$F_{\rm TeV}/F_{\rm X}$ is $\sim~10:1$.
Although TeV $\gamma$-ray emission dominates the X-ray flux in some PWNe
(\cite{Funk2007}, \cite{Kar2009}, \cite{MGH2009}), it is difficult 
to explain such a large ratio with a simple
energy distribution of high energy electrons.

If the origin of TeV $\gamma$-ray emission is inverse Compton scattering 
of cosmic microwave background (CMB) by TeV electrons, 
the same high energy electrons
also emit X-rays by synchrotron radiation. In this case, the ratio of
TeV flux to the X-ray flux depends only on the magnetic field strength. 
The synchrotron emission model of TeV electrons are shown by solid lines in 
Fig.~\ref{fig:wide_spec} for some assumed magnetic fields of 1, 3, and
10~$\mu$G.
Our data corresponds to the magnetic field
of about 1~$\mu$G.
However, this value is much lower than that expected in this region.
The total magnetic field strength in the Galaxy disk is larger than 3~$\mu$G
in the whole area \citep{Beck2001}.
In addition, active star forming region indicates stronger magnetic field.

If a photon energy density of the local radiation field is much 
higher than CMB, Compton scattered TeV $\gamma$-rays exhibit 
larger intensity and might explain our data.
The number of OB stars are $\sim 1000$ in Cyg OB2 region.
Although we could not rule out the possibility
of the strong radiation background, we only consider CMB in this study.

One possibility for resolving the discrepancy is a hadronic origin of
TeV $\gamma$-ray emission (e.g., \cite{Bed2003}). In such a case, X-ray emission 
can be much lower than leptonic case, and a typical strength of magnetic field of PWNe, 
3~$\mu$G, could be acceptable.

Another possibility is an existence of a high energy cut-off of electrons.
The energy of electrons, which is responsible for TeV emission, 
is lower than X-ray emitting electrons by about one order of magnitude.
If the energy distribution of TeV
electrons exhibit strong cut-off above the TeV region, the flux of X-ray
emission is decreased (dashed line in Fig.~\ref{fig:wide_spec}).
\citet{Mat2009} investigated the evolution of the PWNe by comparing 14 samples,
and concluded that the ratio of TeV and X-ray luminosities has a positive correlation 
with the characteristic age of the pulsar. The X-ray luminosity decreases with
the pulsar age because of radiative cooling in the earlier stage 
(by a factor $\sim 10^6$ in $10^5$ yr), 
while the $\gamma$-ray luminosity is constant. Thus,
the ratio becomes larger. 
PSR~J2032$+$4127 is an old pulsar
of 120~kyr \citep{Abdo2010b}. The large
luminosity ratio might be natural for such an evolved PWN.

This cut-off hypothesis also explains the difference in size between TeV and X-ray emissions
as discussed in \citet{Aha2005b}.
X-rays are emitted by young electrons, because higher energy electrons lose their energy
quickly. On the other hand, older electrons radiate TeV $\gamma$-rays.
Such a difference in age could cause the concentration of X-ray emission
near the pulsar and the diffusion of TeV emission.
Thus the size of X-ray emitting region could be much smaller than that of TeV.

\subsection{Other structures}
The best-fit parameters of structure~2 are almost the same as structure~1.
In addition, the X-ray emission size is also comparable to structure~1. 
Although no pulsar is found at the location of structure~2,
the X-ray emission can be explained by PWN in the same manner
as structure~1. If structure~2 is also a PWN, a part of TeV emission
could originate from this source.

The remaining diffuse emission also exhibits the same spectral
shape as structure~1. However, the X-ray flux and the size is much
larger than structure~1, 2. We cannot insist that the remain is related
to TeV emission only by the spectral similarities. 

\section{Summary}

We observed the first unidentified TeV $\gamma$-ray source 
TeV J2032$+$4130 with Suzaku, and detected two structures
of diffuse X-ray emission. The position of the structure 1 is coincident with
the GeV $\gamma$-ray pulsar. We also detected a hint of diffuse emission
extended to whole region of TeV emission.
By estimating the contribution of point sources by Chandra,
we extracted the X-ray spectra of diffuse components.
X-ray and TeV $\gamma$-ray emission
can be explained by electrons with a high energy cut-off above
the TeV region, which could originate in old PWN. 
Such an energy distribution of electrons may lead
smaller X-ray emission size in comparison with TeV emission.

\bigskip

This research made use of data obtained from Data ARchives and Transmission System (DARTS), 
provided by Center for Science-satellite Operation and Data Archives (C-SODA) at ISAS/JAXA.

\end{document}